\documentclass[aps,pre,groupedaddress,twocolumn]{revtex4}
 
\usepackage[utf8]{inputenc}
\usepackage[T1]{fontenc}
\usepackage{ae}
\usepackage{bm}
\usepackage[final]{graphicx}
\usepackage{amsmath,amsfonts,amstext,amssymb,amsbsy,amsopn,amsthm,eucal}

\begin{document}

\title{Continuity equation for probability as a requirement of inference over paths}

\author{Diego González}
\email{dgonzalez@gnm.cl}
\affiliation{Grupo de Nanomateriales, Departamento de Física, Facultad de Ciencias, Universidad de Chile.}

\author{Daniela D\'{i}az}
\affiliation{Facultad de Física, Pontificia Universidad Católica de Chile.}

\author{Sergio Davis}
\email{sdavis@cchen.cl}
\affiliation{Comisión Chilena de Energía Nuclear, Casilla 188-D, Santiago.}

\begin{abstract}
In this work we present the fundamental ideas of inference over paths, and show how this formalism implies 
the continuity equation, which is central for the derivation of the main partial differential equations 
that constitute non-equilibrium statistical mechanics. Equations such as the Liouville equation, Fokker-Planck 
equation, among others can be recovered as particular cases of the continuity equation, under different probability 
fluxes. We derive the continuity equation in its most general form through what we call the \emph{time-slicing 
equation}, which lays down the procedure to go from the representation in terms of a path probability functional 
$\rho[X()]$ to a time-dependent probability density $\rho(x; t)$. The original probability functional $\rho[X()]$ 
can in principle be constructed from different methods of inference; in this work we sketch an application using 
the maximum path entropy or \emph{maximum caliber} principle.
\end{abstract}

\maketitle

\newpage
\section{Introduction}

The study of non-equilibrium systems is an active area of research, both with important applications in 
biology~\cite{Collin2005}, materials science~\cite{Trepagnier2004}, fluid~\cite{Cardy2009} and plasma physics, 
financial modelling and other complex systems, as well as in more fundamental research aiming to understand phenomena such 
as irreversibility and the origin of the second law of Thermodynamics. However, unlike equilibrium statistical mechanics, which 
is a well-established and unified theory that can be derived in its entirety from a single fundamental principle, namely the principle of maximum 
entropy~\cite{Jaynes1957}, non-equilibrium statistical mechanics (NESM)~\cite{DeGroot1984, Zwanzig2001} does not have an 
axiomatic formulation: rather it is presented as a set of partial differential equations (PDEs) and identities, each one with 
restricted validity (for instance, some of them are only valid in Onsager's linear regime).

It is immediate to note that all the different partial differential equations for the time-dependent probability $\rho({\bm x}; t)$ that 
appear in NESM are, in fact continuity equations, of the form 

\begin{equation}
\partial_{t}\rho + \nabla \cdot (\rho \bm{v}) = 0.
\label{eq_continuity}
\end{equation}
Therefore it is possible to obtain the full set of PDEs as particular cases of a continuity equation with different \emph{flow velocities} 
$\bm{v}$. In turn, the continuity equation, which represents local conservation of the probability, is usually obtained via \emph{ad hoc} 
arguments~\cite{Evans2007} involving fluxes in and out of a given surface. This is of course valid, but it is not clear what is the 
range of validity of this kind of derivation for more abstract systems. 

In this work we show that the continuity equation (Eq. \ref{eq_continuity}) is a direct consequence of performing statistical inference 
in the space of paths from a point $A$ to a point $B$. For this result we employ what we call the \emph{time-slicing} equation, a 
rule which extracts information about the probability of different microscopic states on a slice of constant time from an ensemble of 
dynamical paths.

\section{Continuity equation and Non-equilibrium Statistical Mechanics}

A crucial element in Non-equilibrium Statistical Mechanics is the time-dependent probability density of microstates $\rho({\bm x};t)$. This is 
typically obtained by solving a partial differential equation appropriate for the system, for instance such as the Fokker-Plank equation for 
several kinds of Brownian motion models, and the Liouville Equation in the case of time evolution in phase space. All these equations 
can be recast as particular cases of the continuity equation, Eq. \ref{eq_continuity}. Let us write this equation using index notation as 

\begin{equation}
\partial_t \rho(\bm x;t) + \partial_\alpha (\rho(\bm x; t) v^\alpha(\bm x, t)) = 0,
\label{continuity}
\end{equation}
where $\partial_{\alpha}=\partial/\partial x_\alpha$ with $\alpha$=1,2,3, and $\partial_t=\partial/\partial t$. Here the different 
forms of the probability current ${\bm v}(\bm x, t)$ give us the different equations for the probability density $\rho$. As a example, 
consider the case where $v^\alpha({\bm x},t)$ is given by

\begin{equation}
v^\alpha({\bm x},t) = \mu^\alpha({\bm x},t) - \frac{1}{2}\partial_\beta D^{\alpha \beta}({\bm x},t) 
-\frac{1}{2} D^{\alpha \beta}({\bm x},t) \partial_\beta \ln \rho({\bm x};t).
\label{FP-FLUX}
\end{equation}
with $\mu^\alpha$ the components of a drift vector and $D^{\alpha \beta}$ the components of a diffusion tensor. Replacing 
in Eq. \ref{continuity} we recover the Fokker-Plank equation,

\begin{equation}
\partial_t \rho({\bm x}; t) + \partial_\alpha (\rho({\bm x}; t) \mu^\alpha({\bm x},t)) 
-\frac{1}{2}\partial_\alpha\partial_\beta (D^{\alpha \beta}({\bm x},t) \rho({\bm x}; t)) = 0.
\end{equation}

Similarly, the Liouville equation

\begin{equation}
\partial_t \rho({\bm r}, {\bm p}; t) + \big\{\rho, \mathcal{H}\big\} = 0,
\end{equation} 
with $\big\{A,B\big\}$ the Poisson bracket,

\begin{equation}
\big\{A, B\big\} = \sum_{i=1}^{3N} \frac{\partial A}{\partial x_i}\frac{\partial B}{\partial p_i} 
- \frac{\partial A}{\partial p_i}\frac{\partial B}{\partial x_i},
\end{equation}
can be recovered with the choice 

\begin{equation}
{\bm v}({\bm r}, {\bm p}, t) = -\sum_{i=1}^{3N}\hat{e}_i \frac{\partial \mathcal{H}}{\partial x_i} 
+ \sum_{i=3N+1}^{6N}\hat{e}_i \frac{\partial \mathcal{H}}{\partial p_i}.
\end{equation}

It seems that establishing the formal basis for Non-equilibrium Statistical Mechanics requires a solid justification for 
the nature of the continuity equation and its range of applicability. In the following sections we deduce the general continuity 
equation as in Eq. \ref{eq_continuity} purely from considerations of statistical inference over dynamical paths.

\section{Inference over the space of paths and the time-slicing equation}

Dynamical systems (parameterized by time $t$) follow paths which are smooth curves\footnote{Smooth although not 
neccessarily differentiable.}, such that the coordinates at a time $t$ are given by vector functions ${\bm X}(t)$. 
We will assume that there is some uncertainty about the path the system will take when going from a point $A$ to a point $B$, 
and that justifies the use of probabilities. 

Before proceeding, let us make a few remarks about notation. In the following, we will denote a complete path by its function 
${\bm X}()$ without explicitly writing its argument. A functional $G$ of the path ${\bm X}()$ will be denoted by $G[{\bm X}()]$. 
In contrast, a state of the system will be denoted by $\bm x$, in lowercase, so that, for instance, ${\bm x}_0={\bm X}(t_0)$ is the 
condition that the system is in state ${\bm x}_0$ at time $t_0$ when following the path ${\bm X}()$.

\begin{figure}[h!]
\begin{center}
\includegraphics[scale=0.35]{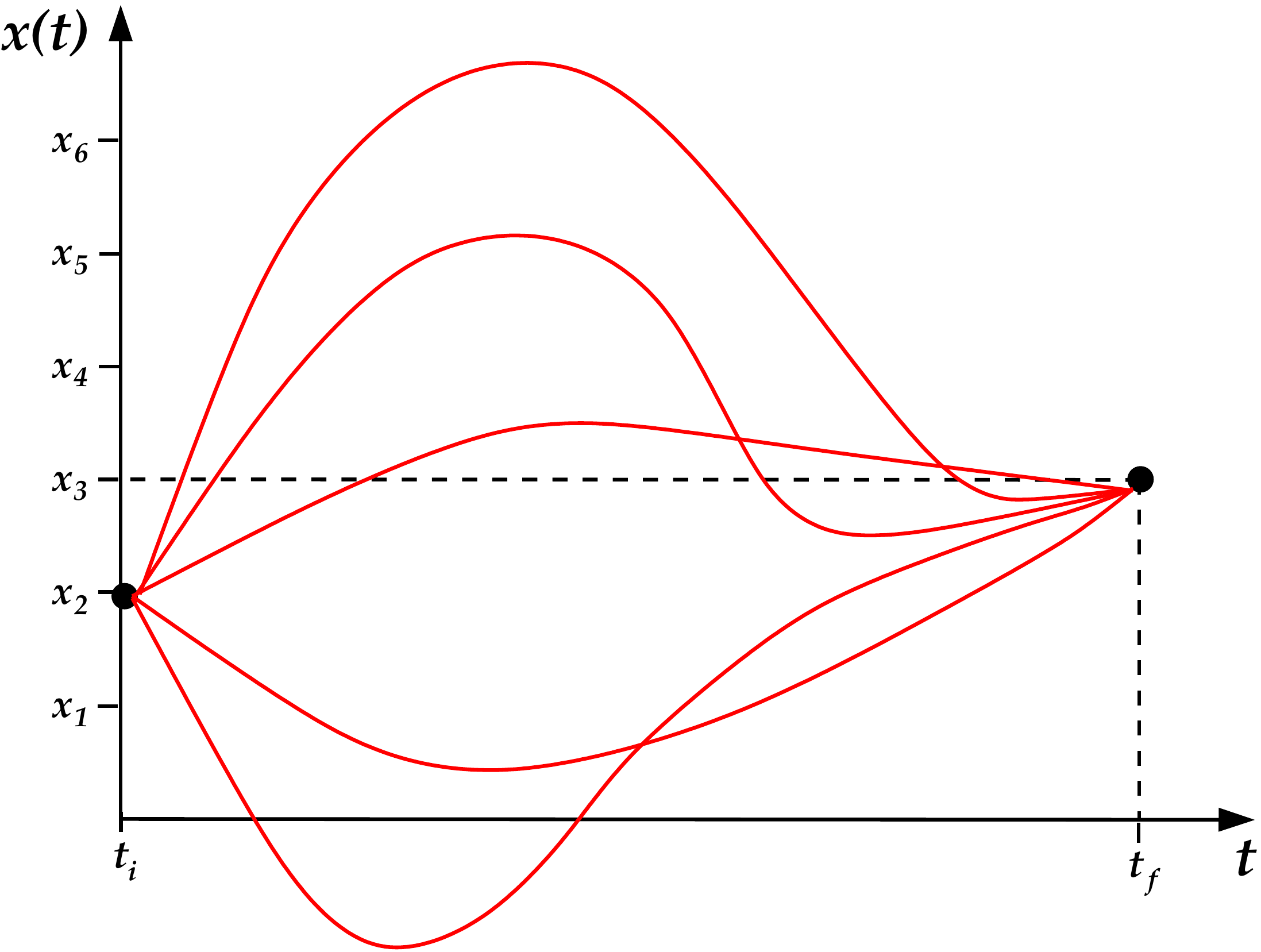}
\end{center}
\caption{Some continuous paths in a path space $\mathbb{X}$ with boundary conditions $X(t_i)=x_2$ and $X(t_f)=x_3$.}
\label{pathspace}
\end{figure}

Imposing fixed boundary conditions defines a space of smooth paths $\mathbb{X}$ such that the actual evolution of a system 
in a given realization of the process is described by some path ${\bm X}() \in \mathbb{X}$. An example of such a path space 
$\mathbb{X}$ in one spatial dimension, where paths have fixed boundary conditions $X(t_i)=x_2$ and $X(t_f)=x_3$, is sketched 
in Fig. \ref{pathspace}.

Now, in order to do inference over dynamical properties we can introduce the concept of a \emph{probability functional} 
$\rho[{\bm X}()]$, which gives the probability \emph{density} assigned to every possible path ${\bm X}() \in \mathbb{X}$. If 
this functional is known, we can in principle estimate any observable $G[{\bm X}()]$ by computing its expectation 

\begin{equation}
\Big<G\Big> = \int_{\mathbb{X}} \mathcal{D}{\bm X}() \rho[{\bm X}()] G[{\bm X}()].
\end{equation}

These expectations are given by path integrals. However, we usually need to determine the expectation of instantaneous 
quantities $g({\bm X}(t))$ which should be estimated using the instantaneous (or time-dependent) probability
density of the states $\rho({\bm x}; t)$, as

\begin{equation}
\Big<g({\bm X}(t))\Big> = \Big<g({\bm x})\Big>_t = \int d{\bm x}\rho({\bm x}; t)g({\bm x}).
\end{equation}

The connection between the two formalisms is given by the representation of $\rho({\bm x}; t)$ as an expectation of a 
Dirac delta functional over the path distribution, 

\begin{equation}
\rho({\bm x}; t) = \Big<\delta({\bm X}(t)-{\bm x})\Big>=\int_\mathbb{X} \mathcal{D}{\bm X}() \rho[{\bm X}()] \delta({\bm X}(t) - {\bm x}).
\label{corte}
\end{equation}

This equality defines the \emph{time-sliced} or instantaneous probability, and we will refer to it as the \emph{time-slicing equation}.
Notice that we employ the same symbol $\rho$ for the probability density of states and for the probability functional of paths, each 
case should be clear in context by the presence of square brackets to signal functional evaluation. 

\section{Properties of the time-slicing equation}

When performing operations on probabilities, a more explicit notation is preferable, in which we use the \emph{probability operator} 
(denoted by the capital letter $P$). The operator $P$ is a binary operator, which takes two assertions (or logical propositions) $A$ 
and $B$ and combines them giving a real, non-negative number $P(A|B)$. This should be read as ``the probability of $A$ being true 
given that $B$ is true''. The probabilities we have employed so far, $\rho({\bm x}; t)$ and $\rho[{\bm Y}()]$, can be written explicitly as 

\begin{eqnarray}
\rho({\bm x}; t) = P({\bm X}(t) = {\bm x}|I), \\
\rho[{\bm Y}()] = P({\bm X}() = {\bm Y}()|I),
\end{eqnarray}
where ${\bm X}()$ represents the real path followed by the system, and $I$ is a \emph{prior} state of knowledge. Using these 
definitions and the rules for probability, we will prove a more general relation,

\begin{equation}
\Big<\delta({\bm X}(t)-{\bm x})G[{\bm X}()]\Big>_I = \rho({\bm x}; t)\Big<G[{\bm X}()]\Big>_{{\bm x}, t},
\label{eq_slice_rule}
\end{equation}
where $\big<G\big>_{x,t}$ is the \emph{time-sliced expectation} of the functional $G$, i.e., the expectation 
that considers only the paths where ${\bm X}(t)={\bm x}$. This may be considered a \emph{general time-slicing 
equation}, which allows to ``slice'' any functional $G$ by taking expectation with the appropriate delta function. 
To see why this relation is true, let us write the expectation in the left-hand side of Eq. \ref{eq_slice_rule} as a path integral, 

\begin{equation}
\Big<\delta({\bm X}(t)-{\bm x})G[{\bm X}()]\Big>_I = \int_\mathbb{X} \mathcal{D}{\bm X}() P({\bm X}()|I)\delta({\bm X}(t)-{\bm x})G[{\bm x}()],
\end{equation}
and recognize the Dirac delta function as the probability of the state at a specific time given the path, that is, 

\begin{equation}
P({\bm X}(t) = {\bm x}|{\bm X}()) = \delta({\bm X}(t)-{\bm x}).
\end{equation}

Invoking Bayes' theorem as

\begin{equation}
P({\bm X}(t)-{\bm x}|{\bm X}()) = \frac{P({\bm X}(t)={\bm x}|I)P({\bm X}()|{\bm X}(t)={\bm x}, I)}{P({\bm X}()|I)}
\end{equation}
and replacing in Eq. \ref{eq_slice_rule}, we finally obtain

\begin{widetext}
\begin{eqnarray}
\Big<\delta({\bm X}(t)-{\bm x})G[{\bm X}()]\Big>_I = \int_\mathbb{X}
\mathcal{D}{\bm X}()P({\bm X}(t)={\bm x}|I)P({\bm X}()|{\bm X}(t)={\bm x},I)G[{\bm X}()] \nonumber \\ 
= \rho({\bm x}; t)\int_\mathbb{X} \mathcal{D}{\bm X}()P({\bm X}()|{\bm X}(t)={\bm x},I)G[{\bm X}()] = \rho({\bm x};t)\Big<G\Big>_{{\bm x}, t}.
\end{eqnarray}
\end{widetext}

Two trivial cases of the identity in Eq. \ref{eq_slice_rule} are: (a) the case with $G=G_0$ (a constant functional) which recovers the 
time-slicing equation, $\big<\delta({\bm X}(t)-{\bm x})\big>=\rho({\bm x};t)$, and (b) the case with $G=G({\bm X}(t))$, for which 
the identity holds immediately because the functional $G$ is constant for all the points where ${\bm X}(t)={\bm x}$ and therefore 
it drops out of the expectation in the left-hand side.

\section{Continuity equation from the time-slicing equation}

Now we will use Eq. \ref{corte} to derive the continuity equation, showing its fundamental role in a theory of dynamical systems. We 
take the partial derivative with respect to time on both sides,

\begin{equation}
\partial_t \rho({\bm x}; t) = \int_\mathbb{X} \mathcal{D}{\bm X}() \rho[{\bm X}()] \partial_t\delta({\bm X}(t)-{\bm x}),
\label{cont_1}
\end{equation}
and use the chain rule on the Dirac delta as

\begin{equation}
\partial_t \delta({\bm X}(t) -{\bm x}) = \frac{\partial}{\partial {\bm X}(t)}\delta({\bm X}(t)-{\bm x})\cdot {\dot {\bm X}}(t).
\end{equation}

Noting that the gradient of any function $\phi({\bm X}(t)-{\bm x})$ with respect to ${\bm X}(t)$ can be expressed as the 
negative gradient with respect to $\bm x$, we obtain 

\begin{equation}
\partial_t \rho({\bm x}; t) = -\frac{\partial}{\partial {\bm x}}\int_\mathbb{X} \mathcal{D}{\bm X}() \rho[{\bm X}()] 
\delta({\bm X}(t)-{\bm x}){\dot {\bm X}}(t),
\end{equation}
which we evaluate using the general time-slicing equation, Eq. \ref{eq_slice_rule}, leading to the continuity equation (Eq. \ref{eq_continuity})

\begin{equation}
\partial_t \rho({\bm x}; t) + \frac{\partial}{\partial {\bm x}}\Big(\rho({\bm x}; t)\big<{\dot {\bm X}}(t)\big>_{{\bm x}, t}\Big) = 0.
\end{equation}
with flow velocity

\begin{equation}
\bm v({\bm x}, t) = \big<{\dot {\bm X}}(t)\big>_{{\bm x}, t}.
\end{equation}

This reveals that the flow velocity is an expectation over paths (in fact, over the \emph{sliced} probability distribution of paths). Because 
of that, it is a functional of $\rho[{\bm X}()]$ and this gives the possibility of different particular forms of the continuity equation for 
different systems, depending on the kind of paths it explores; for instance, the form of $\bm v$ could depend on the physical, 
macroscopic constraints it is subjected to.

\section{Time-slicing in a discrete model}

In order to explain the time-slicing process, let us consider a simple model where we have discretized both space and time. 
In this case, $t \rightarrow t_i=i\Delta t$ and $x \rightarrow x_j=j\Delta x$, with $i=1,\ldots,M$ and $j=1,\ldots,N$. An 
example of this discretization with $N=5$ and $M=6$ is shown in Fig. \ref{pathspace}.

\begin{figure}[h!]
\begin{center}
\includegraphics[scale=0.45]{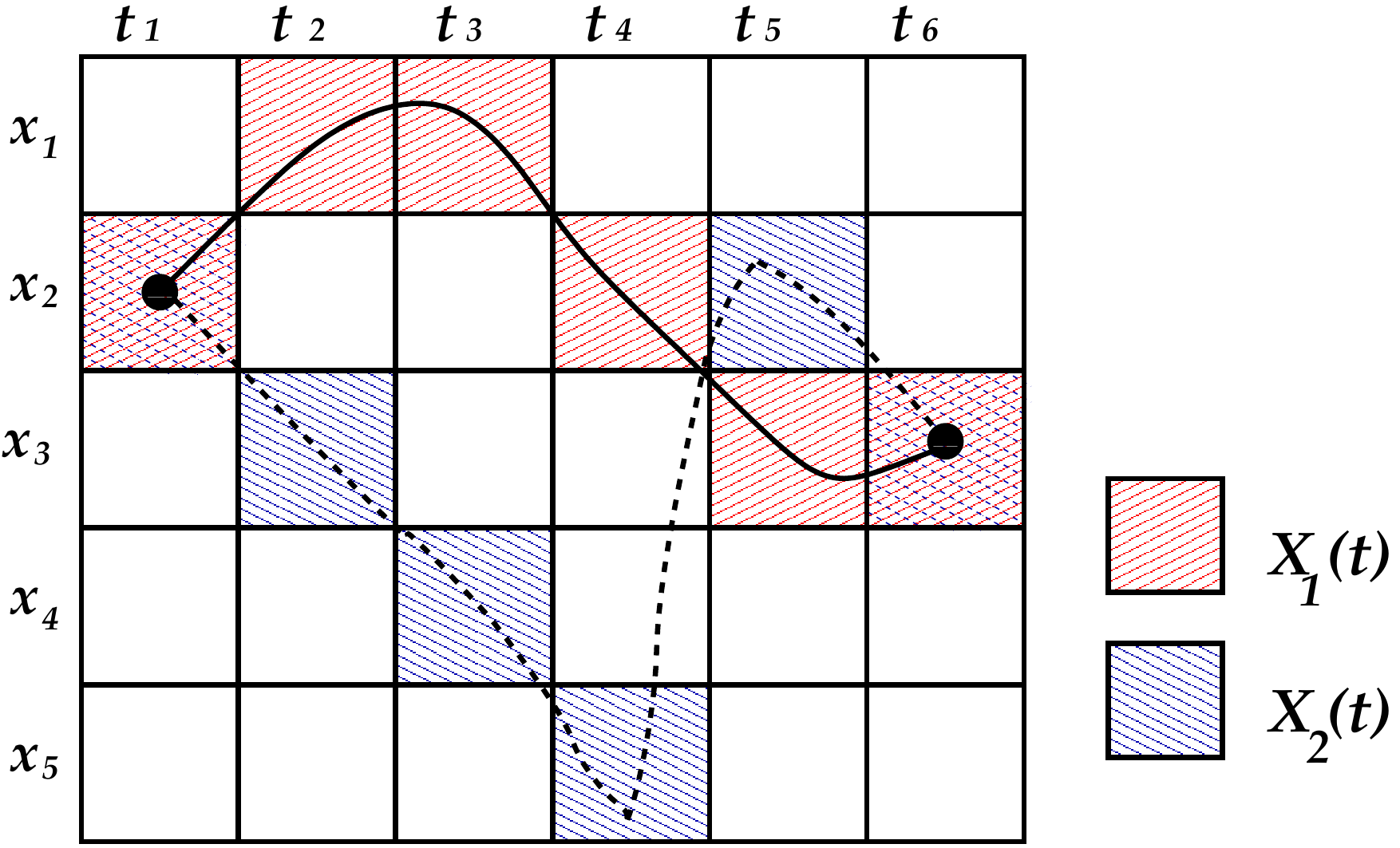}
\end{center}
\caption{Discretized space and time with $N=5$ and $M=6$. Two paths are shown, the solid line represents a smooth path 
while the path depicted with the dotted line is discontinuous.}
\label{grilla}
\end{figure}

Here paths can be represented as sequences of positions, $X=\{X_1, X_2, \ldots, X_M\}$, where $X_i=X(t_i) \in \{x_1, \ldots, x_N\}$. As 
a example, the solid and dotted paths in Fig. \ref{grilla} are given by $\{ x_2, x_1, x_1, x_2,  x_3, x_3\}$ and 
$\{ x_2, x_3, x_4, x_5,  x_2, x_3\}$, respectively. The probability functional
for paths, $\rho[X()]$, reduces to a joint probability of discrete values, $\rho(X)=P(X_1, X_2, \ldots, X_M|I)$.

Once we have proposed a model for the path probability $\rho(X)$, derived from some inference method, we can use 
the time-slicing equation to obtain the probability for the system to be at a discrete position $x_i$ at each time $t_j$, 
which we will denote $\rho_{ij}=P(X_j = x_i|I)$. The connection between $\rho(X)$ and $\rho_{ij}$ is of course given by the 
discrete version of the time-slicing equation,  

\begin{equation}
\sum_X \rho(X) \delta(X_j, x_i) = \rho_{ij},
\label{corte_discreto}
\end{equation}
where $\delta(a, b)$ is the Kronecker delta and the sum is performed over all possible discrete paths $X$. 
The calculation of the state probability $\rho_{ij}$ then reduces to a simple counting of the number of paths $X$ 
crossing the point $x_i$ at the time $t_j$, weighted by the path probability.

\section{Concluding remarks}

We have shown that the idea of performing inference over a path space $\mathbb{X}$ is a promising foundation 
for the study of the dynamics of non-equilibrium systems. Its fundamental role is manifest through the use of what 
we call the time-slicing equation (Eq. \ref{corte}), which connects a probability functional obtained from a variational principle (such 
as the maximum caliber principle) in path space, with a time-dependent probability density for the states. The time-slicing 
equation, which is a definition of a ``slice'' in time consistent with the laws of probability, is a mathematical identity requiring 
no underlying physical principle to be valid.

A direct consequence of the time-slicing equation is the continuity equation for the time-dependent probability density 
(Eq. \ref{eq_continuity}). This reveals the possibility of obtaining different PDEs governing the non-equilibrium statistical mechanics of a 
variety of systems, simply by ``plugging in'' the correct form of the probability current $\big<\dot{X}(t)^b\big>_c$, which is 
dependent on the particular details of each system. We propose that this information must be encoded into the probability functional 
$\rho[X()]$ using a principle such as maximum caliber.

\section{Acknowledgments}

DG and SD thankfully acknowledge funding from FONDECYT grant 1140514. DG acknowledges funding from CONICYT PhD fellowship 21140914.

\bibliography{maxcal}
\bibliographystyle{ieeetr}

\appendix

\section{Construction of a probability functional using the Maximum Caliber principle}

The principle of \textbf{Maximum Caliber}, suggested by Jaynes~\cite{Jaynes1980}, postulates that the most unbiased 
probability distribution of paths is the one that maximizes their Shannon entropy. This entropy of paths is sometimes 
called the ``caliber'' of the system, and is given by the path integral

\begin{equation}
\mathcal{S} = -\int \mathcal{D}{\bm X}() \rho[{\bm X}()] \ln \frac{\rho[{\bm X}()]}{\Pi[{\bm X}()]},
\end{equation}
where $\Pi[{\bm X}()]$ is an invariant measure of paths, which usually is taken as a constant. Under the macroscopic 
constraint on the expectation of an instantaneous function $f$, 

\begin{equation}
\big<f({\bm X}(t), \dot {\bm X}(t), t)\big>=F(t),
\label{t_const}
\end{equation}
for each instant $t$ in an interval $[0, T]$, the normalized probability that maximizes $\mathcal{S}$ is  

\begin{equation}
P({\bm X}()| l()) = \frac{1}{Z[\lambda()]}\exp\Big(-\int_0^T dt\lambda(t)f({\bm X}(t),\dot {\bm X}(t), t)\Big)
\label{eq_p}
\end{equation}
where $Z[\lambda() ]$ is a \emph{partition functional} imposing normalization, and given by

\begin{equation}
Z[\lambda()] = \int \mathcal{D}{\bm X}() \exp\Big(-\int_0^T dt\lambda(t) f({\bm X}(t),\dot {\bm X}(t), t)\Big).
\label{eq_z}
\end{equation}

The maximum caliber path probabilities have interesting properties. For instance, renaming $\mathcal{L}(x, \dot x, t)=\lambda(t)f(x, \dot x, t)$ 
we see that the exponent in Eq. \ref{eq_p} has the form the action of a classical system with Lagrangian $\mathcal{L}$~\cite{Gonzalez2014, Davis2015}. Thus it is useful to rewrite it as 

\begin{equation}
P({\bm X}()| l()) = \frac{1}{Z[\lambda()]}\exp(-A[{\bm X}()]) 
\end{equation}
with 
\begin{equation}
A[{\bm X}()] = \int_0^T dt\mathcal{L}({\bm X}(t),\dot {\bm X}(t); t)
\end{equation}

Eqs. (\ref{eq_p}) and (\ref{eq_z}) give us a tool for making inferences about paths ${\bm X}()$ under known information 
in the form of time-dependent expectation values. 

\begin{figure}[h!]
\begin{center}
\includegraphics[scale=0.35]{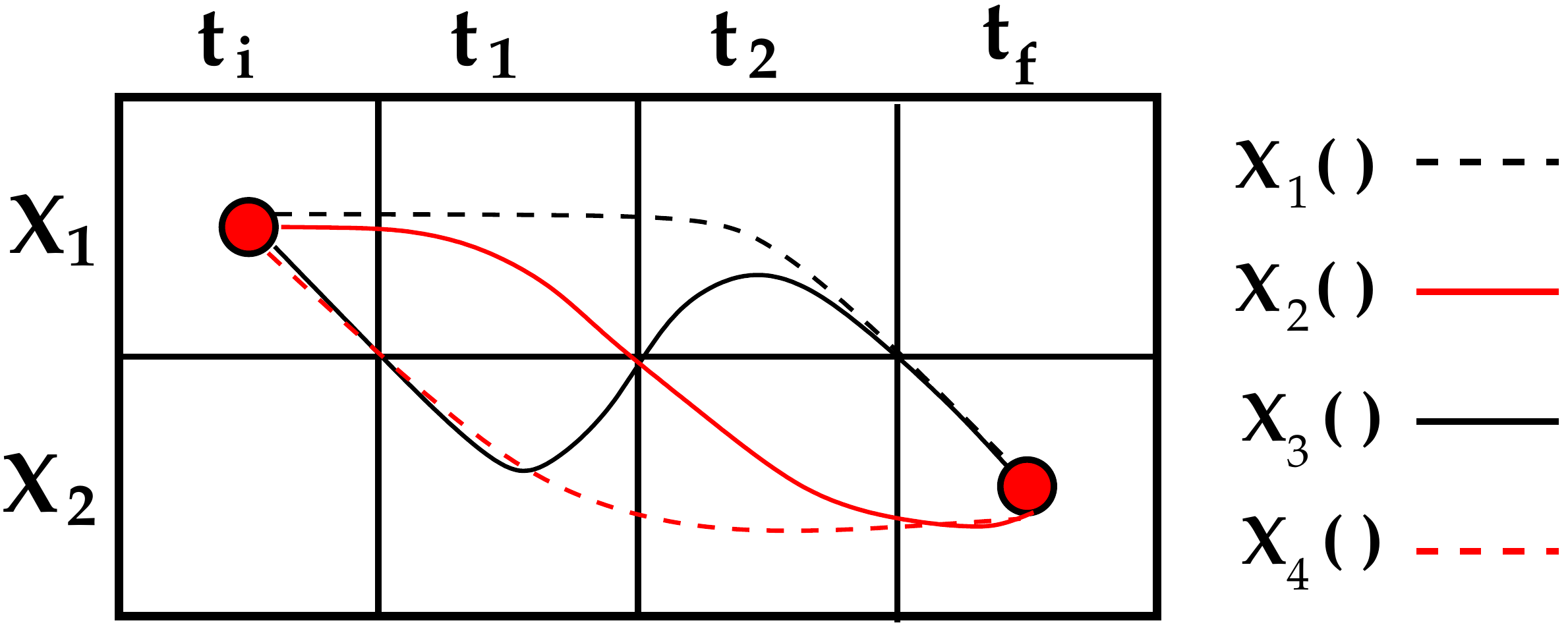}
\end{center}
\caption{Discretized model of $2 \times 4$ with the 4 possibles paths following the boundary conditions.}
\label{2x4grid}
\end{figure}

In order to see how the Maximum Caliber principle and the time-slicing equation work together, consider a 
discretized model with 2 possible positions $x_1$ and $x_2$, and 4 possible times $t_1$, $t_2$, $t_3$ and $t_4$. 
There are 4 allowed paths from $x_1$ in $t_1$ to $x_2$ in $t_4$, described by the vectors $X=(x_1, X_2, X_3, x_2)$
with $X_2, X_3 \in \{x_1, x_2\}$. All these paths are shown in Fig. \ref{2x4grid}.

Let us assign a simple Lagrangian $\mathcal{L}$ to this discretized space, such as $\mathcal{L}(x_k, t_j)=2\delta_{kj}-1$. 
Here the Lagrangian is independent of velocity, just for clarity. Then, the action for each path can be calculated exactly; 
these values are shown in Table \ref{tbl_paths}. 

\begin{table}[b!]
\begin{tabular}{|c|c|c|}
\hline
Path & Action & Probability \\
\hline
$\Gamma_1=\{x_1, x_1, x_1, x_2\}$ & -2 & 0.1050  \\
$\Gamma_2=\{x_1, x_1, x_2, x_2\}$ & 0  & 0.0142  \\
$\Gamma_3=\{x_1, x_2, x_1, x_2\}$ & -4 & 0.7758  \\
$\Gamma_4=\{x_1, x_2, x_2, x_2\}$ & -2 & 0.1050  \\
\hline
\end{tabular}
\caption{The four possible paths joining $x_1$ in $t_1$ and $x_2$ in $t_4$ for a 2x4 discrete model, together with their 
action and probability values.}
\label{tbl_paths}
\end{table}

In the same manner, the partition function can also be computed exactly using Eq. \ref{eq_z}, $Z = \exp(2)+\exp(0)+\exp(4)+\exp(2)\approx$ 
70.37626. With this normalization, the probability for each path is shown in Table \ref{tbl_paths}.
It is also possible to employ the time-slicing equation to obtain the time-dependent probability of visiting each state $x_i$ at the instant 
$t_j$, $\rho_{ij}$, we just compute the sum of the probabilities of each path $\rho(\Gamma)$ passing through a position $x_j$ at time $t_i$, 
where $\rho(\Gamma)$ is given by Eq. \ref{eq_p}. For example, 

\begin{equation}
\rho_{11} = \sum_\Gamma \rho(\Gamma)\delta(X_1,x_1)=\rho(\Gamma_1) + \rho(\Gamma_2) \approx \text{0.1192}.
\end{equation}

The full probability matrix $\rho_{ij}$ are given in Table \ref{tbl_probs}.

\begin{table}
\begin{tabular}{|c|c|}
\hline
Probability & Value \\
\hline
$\rho_{11}$ & 0.1192 \\
$\rho_{21}$ & 0.8808 \\
$\rho_{12}$ & 0.8808 \\
$\rho_{22}$ & 0.1192 \\
\hline
\end{tabular}
\caption{Probability matrix elements for the positions $x_1$ to $x_4$ at each
\label{tbl_probs}
instant $t_1$ and $t_2$.}
\end{table}

We can check that the instantaneous probability is properly normalized for each time, i.e., that 

\begin{equation}
\sum_{j=1}^{2} \rho_{ji} = 1 \;\;\forall i.
\end{equation}

It is important to notice that most probable path (which corresponds to the path with minimum action), $\Gamma_3$, 
is the path that joins the most probable points for each time.

\end{document}